\documentclass[12pt,preprint]{aastex}

\usepackage{graphicx}

\shorttitle{H_2 and H_2 collisions}

\shortauthors{T-G Lee et al.}

\begin{document}

\title{Rotational quenching rate coefficients for H$_2$ in collisions with H$_2$ 
from  2 to 10,000~K}

\author{T.-G.~LEE\altaffilmark{1,2}, N.~BALAKRISHNAN\altaffilmark{3},
R.~C.~FORREY\altaffilmark{4}, \\ P.~C.~STANCIL\altaffilmark{5},
G.~SHAW\altaffilmark{6},
D.~R.~SCHULTZ\altaffilmark{2}, AND G.~ J.~FERLAND\altaffilmark{1}}

\altaffiltext{1}{Department of Physics and Astronomy, University
of Kentucky, Lexington, KY 40506}

\altaffiltext{2}{Physics Division, Oak Ridge National Laboratory,
Oak Ridge, TN 37831}

\altaffiltext{3}{Department of Chemistry, University of
Nevada--Las Vegas, Las Vegas, Nevada 89154}

\altaffiltext{4}{Department of Physics, Penn State University,
Berks Campus, Reading, PA 19610}

\altaffiltext{5}{Department of Physics and Astronomy and Center for
Simulational Physics, University
of Georgia, Athens, GA 30602}

\altaffiltext{6}{Department of Astronomy and Astrophysics, Tata 
Institute of Fundamental Research, Mumbai-400005, India}

\begin{abstract}
Rate coefficients for rotational transitions in H$_2$ induced by
H$_2$ impact are presented. Extensive quantum mechanical
coupled-channel calculations based on a recently published (H$_2$)$_2$
potential energy surface were performed. The potential energy
surface used here is presumed to be more reliable than 
surfaces used in previous work.           
Rotational transition cross sections with
initial levels $J \le$ 8 were computed for collision energies
ranging between 10$^{-4}$ and 2.5 eV, and the corresponding rate
coefficients were calculated for the temperature range 2 $\leq$
$T$ $\leq$ 10,000~K. 
In general, agreement with earlier calculations, which were limited
to 100-6000~K, is good though  discrepancies are 
found at the lowest and highest temperatures. Low-density-limit cooling
functions due to para- and ortho-H$_2$ collisions are obtained 
from the collisional rate coefficients. Implications of 
the new results for non-thermal 
H$_2$ rotational distributions in molecular
regions are also investigated.
\end{abstract}


\section{INTRODUCTION}
Collision-induced energy transfer involving H$_2$ molecules plays
an important role in many areas of astrophysics since
hydrogen is the most abundant molecule in the interstellar
medium (ISM), dominating the mass of both dense and translucent
molecular clouds. Because H$_2$ is a homonuclear molecule
and lacks a dipole moment, its infrared emission is weak and it is
difficult to observe directly. Furthermore, because of the low
mass of the hydrogen atom, the rotational levels of molecular
hydrogen are widely spaced
and require relatively high temperatures, $T \geq$ 100 K, to excite
even the lowest levels appreciably via thermal collisions. 
Nevertheless, with the current
sensitivity of detectors deployed on space-based
observatories, such as the {\it Spitzer Space Telescope} and the
former {\it Infrared
Space Observatory (ISO)}, the detection and mapping of the pure
rotational lines of H$_2$ in a significant sample of galaxies is
now possible \citep{Neu06,Fue99}. These observations, as well
as others from the ultraviolet \citep{mey01,boi05,fra05,fra07} to the 
near infrared \citep{bur98,mcc99,all05} probe shock-induced
heating and cooling and UV and x-ray irradiation of gas in the ISM of galactic and
extragalactic sources and star-forming regions.

Being the simplest diatom-diatom collision system and one in which
all constituents are quantum mechanically indistinguishable, early
work, notably by Green and coworkers, carried out a succession of
theoretical studies on rotational relaxation in H$_2$+H$_2$
collisions \citep{Gre75, RGR77, RRG78, GRR78}. 
The calculation of the relaxation cross sections involve the quantum
mechanical scattering of the heavy particles on a potential
energy surface (PES) which is computed using quantum chemistry techniques,
accounting for the electronic motion for all relevant nuclear   
configurations of the constituent colliders.
The availability of more realistic (H$_2$)$_2$ potential surfaces 
has enabled considerable progress in obtaining a more
reliable set of state-to-state collisional cross sections and
rate coefficients. To date, the most comprehensive
study has been made by \citet{Flo98} and \citet{Flo98a}. 
Within the rigid-rotor
approximation, Flower \& Roueff have made use of 
the (H$_2$)$_2$ interaction potential of \citet{sch88} 
 in a  quantal coupled-channel
method to determine the rate coefficients for rotational
transitions in H$_2$ + H$_2$ collisions.
Rotational levels {\it J} $\le$ 8 and kinetic temperatures $T \le$
1000 K were employed and the results were compared with those reported by Danby,
Flower, \& Monteiro (1987) based on an older potential surface.

Accurate determinations of rate coefficients for state-to-state transitions in
small molecular systems such as the H$_2$ molecule require a quantum 
mechanical description of the scattering process. A full quantum calculation 
of rovibrational transitions in the H$_2$-H$_2$ system is computationally 
challenging. So far,
only a few studies have been reported that include both rotational and 
vibrational degrees of freedom of the H$_2$ molecules.
Pogrebnya and Clary (2002) reported vibrational relaxation of H$_2$ in 
collisions with H$_2$ using a coupled-states approximation implemented within 
a time-independent quantum mechanical approach. Panda et al. (2007) and 
Otto et al. (2008)
employed a time-dependent quantum mechanical approach and reported
rotational transition cross sections in ortho-H$_2$+para-H$_2$ and para-H$_2$
collisions. More recently, Quemener et al. (2008) 
reported a full-dimensional quantum scattering
calculation of rotational and vibrational transitions in the H$_2$-H$_2$
system that does not involve any dynamics approximation. All these calculations adopted
the six-dimensional H$_4$ PES developed by Boothroyd et al. (2002). While
this PES is useful for benchmarking full-dimensional
quantum scattering codes for the H$_2$-H$_2$ system, 
comparisons with experimental data have shown \citep{lee06, Otto08} 
that the PES predicts 
rotational transition rate coefficients between the
 $J=0$ and $J=2$ levels that are too small
and that it is not appropriate for accurate determination
of rotational transition rate constants in the H$_2$-H$_2$ system. 

Since full-dimensional quantum calculations of four-atom systems are computationally
demanding, especially for excited rotational levels, the rigid-rotor approximation
is often employed. The adequacy of the rigid-rotor approximation for pure rotational
transitions in H$_2$-H$_2$ collisions was recently 
demonstrated by Otto et al. (2008)
who reported cross sections for rotational excitations in para-H$_2$ using the
rigid-rotor model and a full-dimensional quantum calculation. While these 
calculations have been performed using the BMKP PES, we believe that 
the conclusions are valid for the H$_2$-H$_2$ system in general.

The accuracy of collisional data for astrophysical applications is limited
by the uncertainty in the PES and the approximations employed in the dynamics 
calculations.
It is a non-trivial task to judge the impact of the
uncertainties in the collisional data, associated with the choice
of the PES. Accurate determination of the potential energy
in the interaction region
to the requisite level of accuracy of $\sim$ 10$^{-3}$ hartrees
(or $\sim$10~K) remains a challenge, as mentioned above. 
This is especially important at low temperatures
where small uncertainties in the interaction potential translate into
larger errors in the cross sections and rate constants. A further difficulty arises
from fitting a limited number of ab initio potential energy points to obtain
a surface over all configurations which are sampled in the scattering
calculations. Unphysical behavior in the fitted PES can result, 
particularly when bridging explicitly calculated energies at 
intermediate  internuclear
separations to long-range asymptotic forms.

Recently, a new and  
improved (H$_2$)$_2$ rigid-rotor PES was computed by \citet{DJ2000}. The accuracy
of this PES for rotational transitions in H$_2$ has recently been demonstrated by
comparing computed rate coefficients for $J=0\to 2$ rotational excitation
against experimental results \citep{Mate,lee06,Otto08}.  Thus, we believe that the 
new PES by Diep \& Johnson (2000) could be employed to provide reliable values of 
rate constants for rotational transitions in the H$_2$+H$_2$ system.
Since the H$_2$+H$_2$ collision system is of
astrophysical significance, it is crucial to establish whether the
cross sections and rate coefficients for rotational energy transfer based on
this PES are in agreement with the earlier data.
Therefore, we adopted this PES in the present quantum mechanical
close-coupling calculations to obtain the rotational transition
cross sections with initial levels $J \le$ 8 for collision
energies ranging from 10$^{-4}$ to 2.5 eV. The corresponding rate
coefficients were computed for the temperature range of 2 
$\leq$ $T$ $\leq$ 10,000~K. The present rate coefficients are
compared with the results of \citet{Flo98} and
\citet{Flo98a}, which are to the
best of our knowledge, the only comprehensive calculations
available for rotational de-excitation rate coefficients. 
We also present H$_2$-H$_2$ cooling functions in the low-density
limit and test the new rate coefficients in simulations of 
UV irradiated molecular gas. Atomic
units are used throughout unless otherwise specified,
i.e., distances are in bohrs (a$_o$) and energy in hartrees
($E_h$).
Recall that 1 a$_o$ = 0.529177 \AA, while 1 $E_h$ = 27.2114 eV
= 219474.635156 cm$^{-1}$ = 627.51 kcal/mole.

\section{COMPUTATIONAL DETAILS}
We carried out quantal coupled-channel calculations for collisions
of H$_2$ with H$_2$ using a fixed bond length of 1.449 a$_o$
(0.7668 \AA). The rigid-rotor H$_2$ target and projectile energy
levels were calculated using a rotational constant of {\it B} =
60.853 cm$^{-1}$. To solve the coupled-channel equations, we used
the hybrid modified log-derivative Airy propagator (Alexander \&
Manolopoulos 1987) in the general purpose non-reactive scattering
code MOLSCAT developed by \citet{hut94}. The
log-derivative matrix is propagated to large intermolecular
separations where the numerical results are matched to the known
asymptotic solutions to extract the physical scattering matrix.
This procedure is carried out for each partial wave.
We have checked that the total integral cross sections 
are converged with respect to the number of partial waves,
as well as the asymptotic matching radius (e.g., $R=$ 40 a$_o$) for
all channels included in the calculations.

In addition to the partial wave convergence, we have checked that
the results are optimized with respect to other parameters
that enter into the scattering calculations. In particular, the
parameters used for the analytical expression for  the Diep \&
Johnson PES,
\begin{equation}
V(R,\theta_1, \theta_2, \phi_{12}) = \sum_{l_1,l_2,l}
V_{l_1,l_2,l}(R)G_{l_1,l_2,l}(\theta_1, \theta_2, \phi_{12}),
\end{equation}
where $V_{l_1,l_2,l}(R)$ are radial expansion coefficients and
$G_{l_1,l_2,l}$($\theta_1$, $\theta_2$, $\phi_{12}$) are
bispherical harmonics. The angles $\theta_1$, $\theta_2$ denote the two
in-plane angles and $\phi_{12}$ is the relative torsional angle. We
used 10 quadrature points each for integration along angular
coordinates $\theta_1,~\theta_2$, and $\phi_{12}$. From the Diep
\& Johnson fit to their numerical PES, the expansion
coefficients $V_{l_1,l_2,l}(R)$ are determined.  They noted
that only the $V_{0,0,0}(R)$, $V_{2,0,2}(R)$, $V_{0,2,2}(R)$ and
$V_{2,2,4}(R)$ terms make significant contribution to the
potential.

In Table \ref{tab1}, we show a comparison between the PES 
adopted by Flower \& Roueff,
the \citet{sch88} potential, and the Diep \& Johnson (H$_2$)$_2$
PES for linear [$\theta_1$, $\theta_2$,
$\phi_{12}$]=[0, 0, 0] and parallel  [$\theta_1$, $\theta_2$,
$\phi_{12}$]=[$\pi$/2,$\pi$/2,0] configurations, respectively.
Although the two potentials show close agreement for both
configurations, the analytic PES of \citet{sch88}
for (H$_2$)$_2$ is known to be accurate only for pairs of hydrogen
molecules with intermolecular separations not greater than $\sim
5$ a$_o$. The small discrepancy may be attributed to the improved
accuracy in the Diep \& Johnson PES which incorporates the correct long-range 
behavior.

The present calculations consider
the hydrogen molecules to be indistinguishable. Symmetric basis
sets were used for para-para and ortho-ortho collisions, whereas,
for para-ortho collisions, asymmetric basis sets were chosen based
on the order of the energy spectrum. The scattering cross sections
for rotational transitions were computed for collision energies
ranging between 10$^{-4}$ and 2.5 eV. 
Four de-excitation cases with
$\Delta J_1=-2$ were considered (i.e., para-para and para-ortho; 
ortho-para and ortho-ortho). 
Note that a sufficiently large, but truncated basis set has been
used to optimize the computation time and minimize the loss of
numerical accuracy (i.e., to $\lesssim5\%$) in the scattering cross
section calculations. Further details about the calculations can be
found in \citet{lee06}.

Rate coefficients for state-to-state rotational transitions were
obtained by averaging the appropriate cross sections over a
Boltzmann distribution of  the relative kinetic energy $E_k$ of the H$_2$ molecules
at a given temperature $T$
\begin{equation}
k_{J_1 J_2 \rightarrow J_1' J_2'}(T)= {{G} \over
{(1+\delta_{J_1J_2})(1+\delta_{J_1^\prime J_2^\prime})} }
\int_{0}^{\infty}dE_k\sigma_{J_1 J_2 \rightarrow J_1'
J_2'}(E_k)E_k e^{(-\beta E_k)},
\label{eq1}
\end{equation}
where $G = \sqrt{\frac{8}{\mu \pi \beta}}\beta^2$,
$\beta =(k_B T)^{-1}$ with $k_B$ being the Boltzmann constant,
and we have adopted the definition of the cross section for
indistinguishable particles as given by \citet{Gre75}.
This definition differs from that adopted by Flower \& Roueff
which was given by \citet{Zar74}. As discussed by
\citet{Danby87}, for single rotational excitation transitions,
the cross section defined by \citet{Gre75} must be divided by
two if $J_1=J_2$ or $J_1^\prime=J_2^\prime$ to prevent
double counting in the determination of the production rate
of $J_1^\prime$ or $J_2^\prime$. 
The factor $(1+\delta_{J_1J_2})(1+\delta_{J_1^\prime J_2^\prime})$ 
in the denominator of eq. (\ref{eq1}) accounts for it.

\section{RESULTS}

\subsection{Comparison of Rate Coefficients}

Fig.~1 shows a comparison between the present theoretical results
and the experimental rate coefficients of \citet{Mate}
for the (0,0)$\rightarrow$(2,0) excitation
transition between 50 and 300~K, where we have used the notation
($J_1,J_2$).
The experimental and current theoretical rate coefficients are in good
agreement as was shown previously in Lee et al. (2006). The theoretical
results obtained by \citet{Flo98}, which used the older PES of Schwenke,
also shows good agreement with the measured and new calculated rate
coefficients. 

Figures 2(a) and 2(b) show the rotational de-excitation rate
coefficients $\Delta J_1 = -2$, $\Delta J_2 = 0$ for
para-para and ortho-ortho collisions,
respectively. For the para-para case, the current rate coefficients for
most transitions are found to be similar to
those obtained by \citet{Flo98} and \citet{Flo98a}, 
but for the important  (2,0)$\rightarrow$(0,0) 
transition, Flower \& Roueff's result is $\sim$10\% smaller for
temperatures less than 1000~K\footnote{We note that the
rate coefficients from \citet{Flo98} are limited to temperatures between 100
and 1000~K, while those given in \citet{Flo98a} extend from
1000 to 4500~K, but only for para-H$_2$ colliders. We supplement 
these in Figs. 2 and 3, and in the discussion
throughout, with additional data given
between 1500 and 6000~K on David Flower's website at 
ccp7.dur.ac.uk/cooling\_by\_h2/node1.html.}.
On the other hand, (8,0)$\rightarrow$(6,0) transition rate coefficients
from \citet{Flo98} and \citet{Flo98a} are larger compared to 
the current results and the agreement appears to deteriorate
as the temperature increases above $\sim$400~K. The
(4,0)$\rightarrow$(2,0) transition, however, is in excellent 
agreement, but only below $\sim$1000~K. Likewise, for the
ortho-ortho transitions, the rate coefficients of Flower \& Roueff
are similar to the current results, except at the highest temperatures.

Figures 3(a) and 3(b) show the rotational de-excitation rate
coefficients for para-ortho collisions. Reasonable agreement
was observed, except for the (8,1)$\rightarrow$(6,1), where
the results of Flower \& Roueff tend to be significantly larger than
the present calculations. This behavior is essentially
identical to that seen for the (8,0)$\rightarrow$(6,0) transition shown in
Fig.~2(a). Discrepancies are also noted for all of the transitions
at the highest temperatures.

\subsection{Rate Coefficient Data and Fitting}

Numerical values for the rate coefficients are given in Table 2.
We have also fitted the computed rate coefficients to an analytical form
similar to that introduced by \citet{leb99}. The adopted form used here is
\begin{equation}
k_{J_1 J_2 \rightarrow J_1' J_2'}(T)= 10^{(a + b/t_1 + c/t^2)}
   + 10^{[e + (f/t_2)^h]}
\end{equation}
where $t=T/1000 + \delta t$, $t_1=d\times T/1000 + \delta t$, 
and $t_2=g \times T/1000 
+ \delta t$, with $\delta t = 1$. The fit coefficients $a$ through $h$,
a Fortran routine to return the rate coefficient $k$ following
input of the temperature $T$ and the set of initial and final quantum
numbers, the numerical data for the rate coefficients, and the numerical
data for the cross sections can all be obtained at the website:
www.physast.uga.edu/ugamop/. The fits are valid between 2 and 10,000~K
and were performed for exothermic, $\Delta J_1 = -2$, transitions.
Rate coefficients for endothermic transitions can be obtained by
detailed balance which is implemented in the Fortran routine.

Because the accuracy of the PESs for values of $l_1$, $l_2$, and $l\geq 4$
(see eq. 1) is uncertain, we did not explicitly compute the rate
coefficients for $\Delta J_1 < -2$ transitions, except for two 
$\Delta J_1 = -4 $ cases: (4,0)$\rightarrow$(0,0) and
(5,1)$\rightarrow$(1,1). The rate coefficients for these processes
are typically a factor of $\sim$50 smaller than the $\Delta J_1 = -2$
rate coefficients, for the same initial state. Transitions with
$\Delta J_1 < -4$ are expected to have even smaller rate coefficients. While these two
rate coefficients have been fitted, we estimate all other $\Delta J_1
< -2$ transitions with a version of the Modified Exponential Energy
Gap Law (MEGL) \citep{ste91}. With the MEGL, the rate coefficients
for $\Delta J_1 < -2$ transitions are obtained by scaling the
$\Delta J_1= -2$ rate coefficients with the following relation
\begin{equation}
k_{J_1 J_2 \rightarrow J_1' J_2'}(T) = k_{J_1 J_2 \rightarrow J_1-2, J_2'}(T)
  {{2(J_1-2) + 1}\over{2J_1'+1}}\exp(-\beta(T)\Delta/T)
  \biggl[{{1+\alpha\Delta/T}\over{1+\alpha\gamma\Delta/T}}\biggr]^2
\end{equation}  
where $\Delta=[E(J_1-2)-E(J_1')]/k_B$ is the difference in rotational
binding energies in K and $\alpha=2$ is usually adopted.  We find 
that the (4,0)$\rightarrow$(0,0) transition is fit with
$\gamma=2/3$ and $\beta(T)/T=1/300$, the latter being nearly temperature independent.

\section{Astrophysical Applications}

In order to investigate the effect of the new rate coefficients in 
astrophysical environments, we show in Figure 4 
low-density-limit cooling functions given by
\begin{equation}
{{\Lambda(T)}\over{n_{\rm H_2}(J_1) n_{\rm H_2}(J_2)}} =  k_{J_1 J_2 \rightarrow J'_1 J_2}(T) h\nu_{J'_1 J_1},
\end{equation}
where $\Lambda(T)$ is the cooling rate in erg cm$^{-3}$ s$^{-1}$, 
$n_{\rm H_2}(J)$ is the number density of H$_2$ in rotational state $J$, and
$h\nu_{J'_1 J_1}$ is the emitted photon energy. Here we consider 
transitions among the lowest four rotational states and plot the
cooling rate due to para-H$_2$ collisions on H$_2$($J_1=0,1$) and
the total cooling rate for a statistical distribution of ortho- and
para-H$_2$, i.e. ortho:para=3:1, following the prescription given
by \citet{Glo08}. The new cooling functions are compared to those obtained
by \citet{Glo08}, based on the H$_2$-H$_2$ rate coefficients of
Flower \& Roueff, which are indicated to be valid between $100 < T < 6000$~K.
The two sets of cooling functions are generally in agreement to
within $\sim$20\% for $T<2000$~K including cooling due to ortho-H$_2$
collisions (not shown). The level of agreement reflects differences
in the collisional rate coefficients shown in Figs. 2 and 3. 
However, above $\sim$2000~K the cooling functions of \citet{Glo08} are seen to
depart significantly from the current results which is due to the
divergence in the two sets of rate coefficients at the higher temperatures. 

In a further test, we include the 
full set of H$_2$-H$_2$ rate coefficients  in a model of 
a molecular photodissociation region using the
spectral simulation code Cloudy \citep{fer98}.
 Model V3 of the PDR-Code
Comparison Study \citep{rol07} is adopted where the total density is
taken to be 10$^{5.5}$ cm$^{-3}$ and the intensity of the  incident far ultraviolet
(FUV) radiation field is 10 times the standard \citet{dra78}
FUV field. Other model parameters of the molecular region are given in
\citet{rol07}. A nearly complete description of the electronic, vibrational,
and rotational  radiative
and collisional properties of the H$_2$ molecule
was included as described by \citet{sha05}. Figure 5a gives the H and H$_2$
number densities and the temperature structure of the cloud as a function
of depth. 

In Figure 5b, the effect of the new H$_2$-H$_2$ rate coefficients
are shown by plotting the ratio of the rotational populations $n(v=0,J)$ 
determined from two separate models. In the numerator, the standard set 
of H$_2$-H$_2$ rate coefficient fits 
given in \citet{leb99} are adopted, while in the denominator
the current H$_2$-H$_2$ rate coefficients are used. H$_2$-H and H$_2$-He
collisions are also included and taken from the recent calculations
of \citet{Wra07} and \citet{Lee08} and used in both models.
Since the rate coefficients from the two calculations are in general
agreement, most levels do not experience a significant change in population. However,
differences are seen for cloud depths greater than $\sim 10^{10}$
cm which corresponds to an increase in the total H$_2$ abundance and
decreases in the H abundance and gas temperature. The abundance changes maximize
the effect of H$_2$ colliders reducing the importance of H collisions, though
He collisions are still important. As such, while the discrepancies
in the H$_2$-H$_2$ rate coefficients are maximum above $\sim$1000~K, their
effects at these high temperatures will be masked by the more 
important roles of H and He collisions.
However, as the temperature decreases below 100~K and H$_2$ collisions
become more important, the discrepancies
are related to the fact that the 
\citet{leb99} fits were not intended to be valid below 100~K.
The largest differences are for the $J=3$ and $J=4$ populations, $\sim$50\% and
$\sim$20\%, respectively. These differences are related to the increase
in the rate coefficients to populate the levels as shown in Figs. 3a and 2a,
respectively, for temperatures less than 200~K.
As the temperature falls further, below $\sim$20~K, 
discrepancies in the $J=2$ and $J=5$ populations also become significant
due to the fact that the current rate coefficients reach a minimum
near this temperature, while the fits of \citet{leb99} go to
constant values.

\section{Summary}

We have performed an extensive quantum mechanical
coupled-channel calculation, based on a recent and presumably more reliable
(H$_2$)$_2$ potential energy surface, to obtain rotational
transition cross sections with initial levels $J \le$ 8. Collision
energies ranging between 10$^{-4}$ and 2.5 eV were considered. We
have computed the corresponding rate coefficients for the
temperature range from 2 to 10,000~K and have compared our results
with the previous work of \citet{Flo98} and
\citet{Flo98a}, which were based on an older potential energy
surface. We conclude that in some cases, the improvement made in
the new potential energy surface and larger number of basis
functions used in the present scattering calculations led to significant
changes in the resulting state-to-state cross sections and rate
coefficients. 
Implementation of the new H$_2$-H$_2$ rate coefficients
in calculations of the low-density-limit cooling function and in
the rotational level populations in a molecular photodissociation region
result in some differences compared to the use of the
rate coefficients from Flower \& Roueff.
Complete tabulations of rate coefficients are available
electronically upon request, on the web at
www.physast.uga.edu/ugamop, and from the ApJ electronic tables.

\acknowledgments

We acknowledge support from NASA grant NNG05GD81G
(TGL, GJF), the
Spitzer Space Telescope Theoretical Research Program (TGL, RCF, GJF),  
NSF grant PHY-0555565 (NB), NASA 
grant NNG06GC94G (NB), NSF grant PHY-0554794 (RCF), and
NSF grant AST-0607733 (PCS). We thank the 
NSF funded Institute for 
Theoretical Atomic, Molecular, and Optical Physics at
the Harvard-Smithsonian Center for Astrophysics for travel
support.

\clearpage

\onecolumn

\begin{figure}
\epsscale{1.0} \plotone{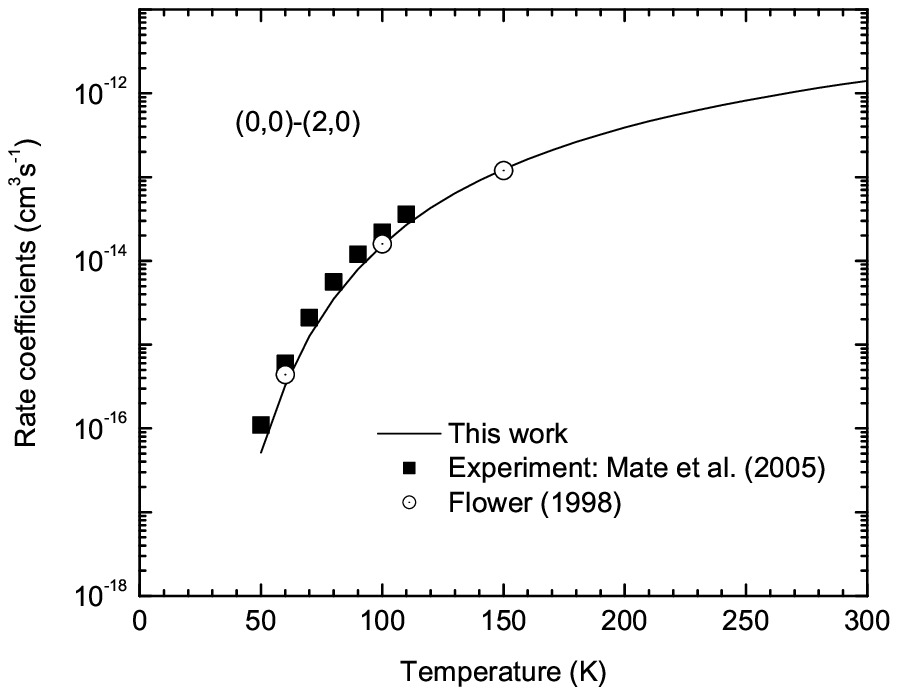} 
\caption{Rate coefficients for
(0,0)$\rightarrow$(2,0) rotational excitation in indistinguishable
H$_2$+H$_2$ collisions as a function of temperature. The solid 
curve denotes the present result using the \citet{DJ2000} PES.
The filled squares are the experimental data from \citet{Mate}.
The open circles are the theoretical results of \citet{Flo98}.
\label{fig1}}
\end{figure}

\begin{figure}
\epsscale{1.10}\plotone{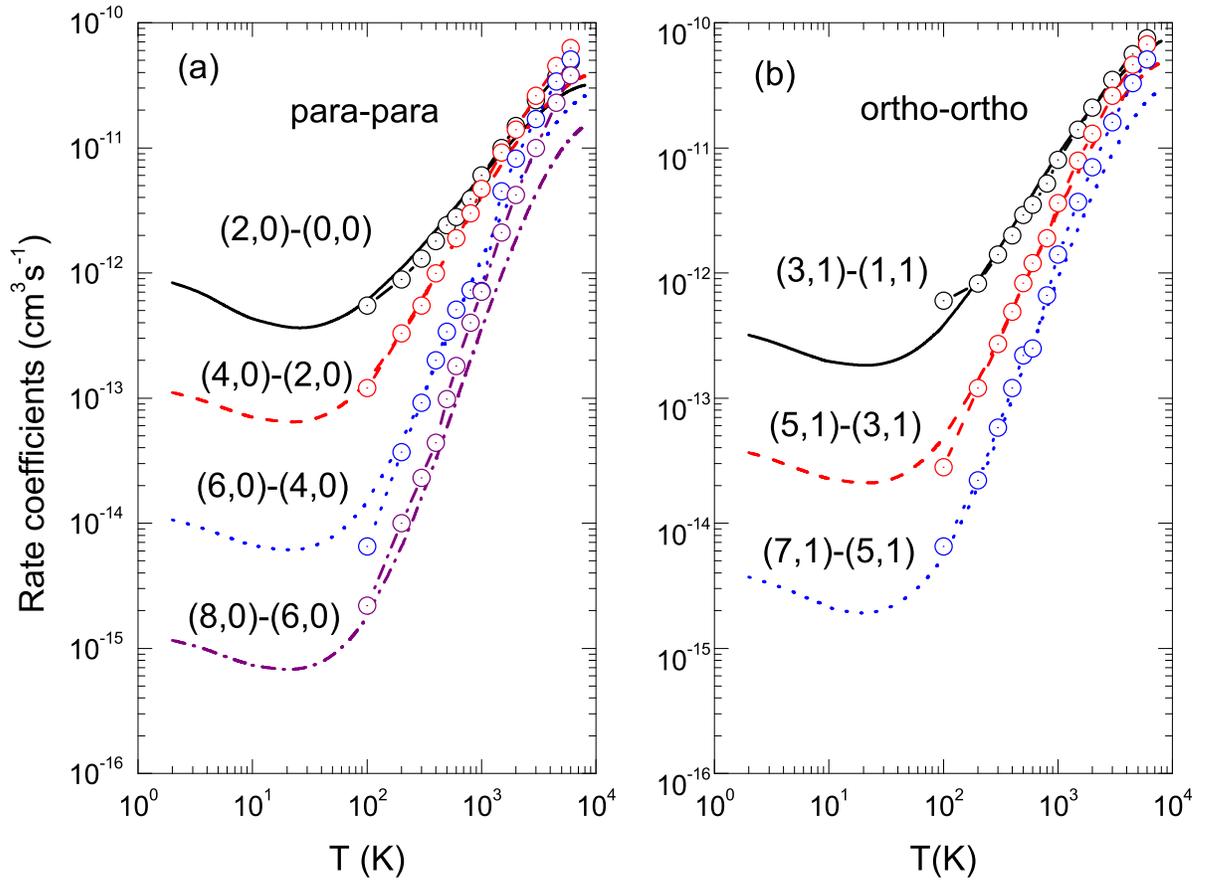} \caption{Temperature dependence
of rotational state-resolved rate coefficients for
H$_2$+H$_2$ collisions. (a) Para-para and (b)
ortho-ortho collisions. Circles denote theoretical results from 
\citet{Flo98} and \citet{Flo98a}. 
\label{fig2}}
\end{figure}

\begin{figure}
\epsscale{1.10}\plotone{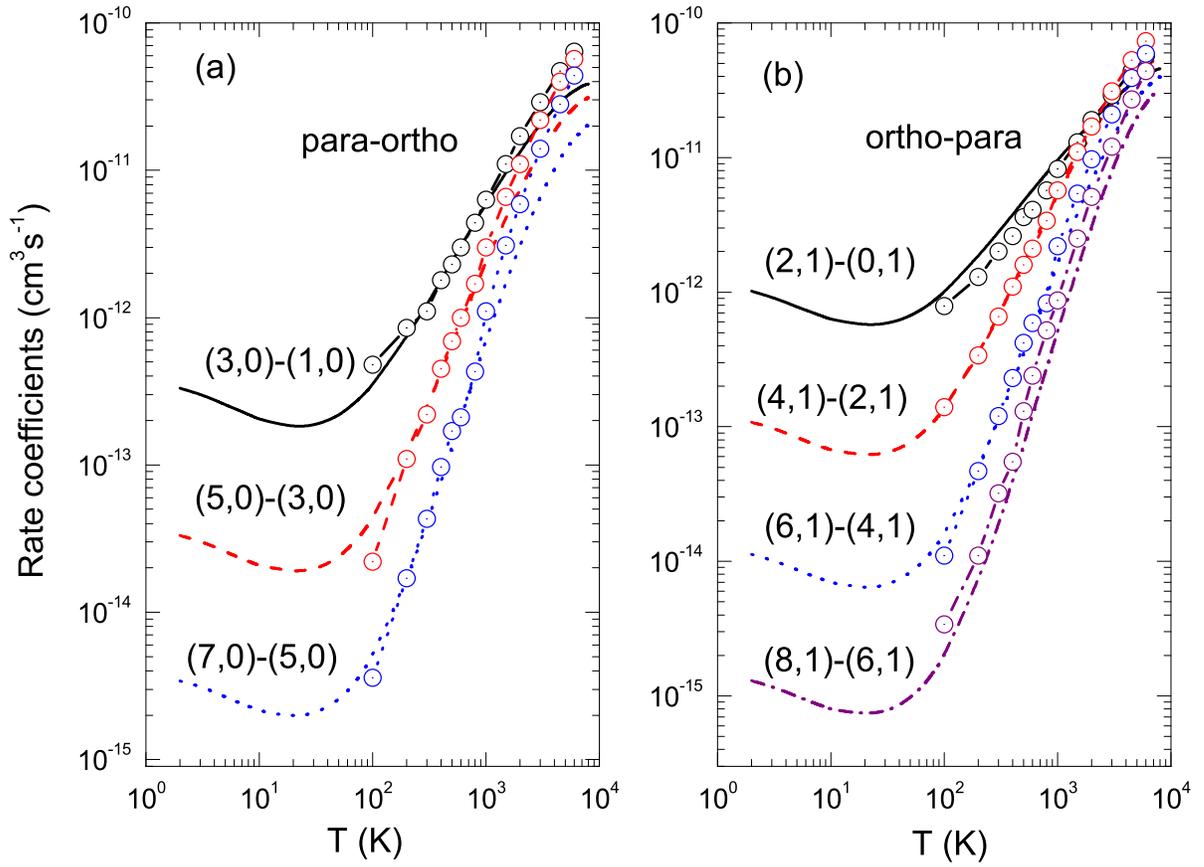} \caption{Same as in Fig.~2, but
for para-ortho collisions. 
\label{fig3}}
\end{figure}

\begin{figure}
\epsscale{1.10}\plotone{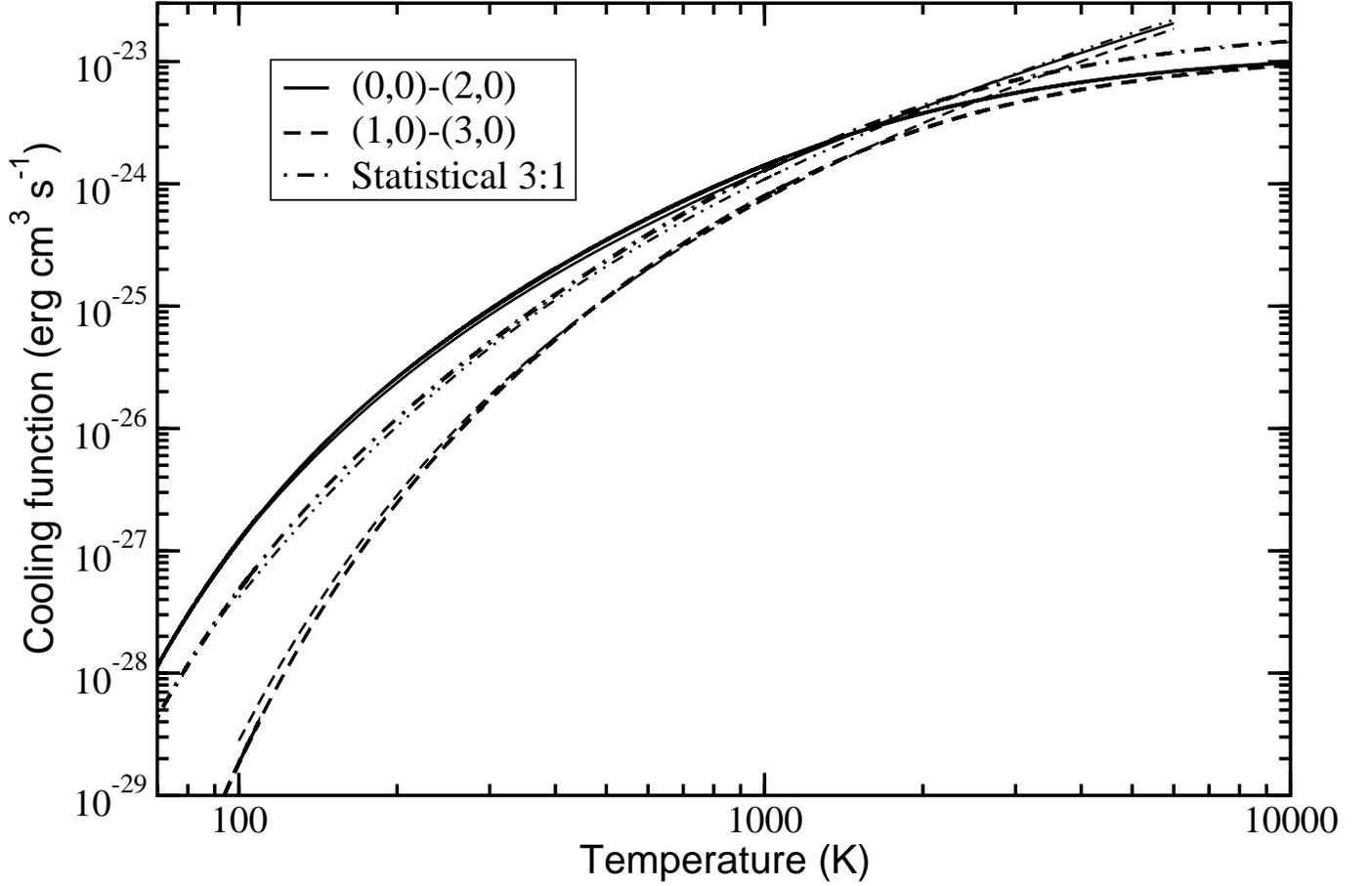} \caption{Low-density-limit
cooling function of H$_2$ due to H$_2$+H$_2$ collisions. Thick lines
show the present results. Thin lines are from \citet{Glo08} based
on the rate coefficients of Flower \& Roueff. Solid, para-para; dash,
para-ortho; and dot-dash, statistical ortho-para population.
\label{fig4}}
\end{figure}

\begin{figure}
\epsscale{1.0}\plotone{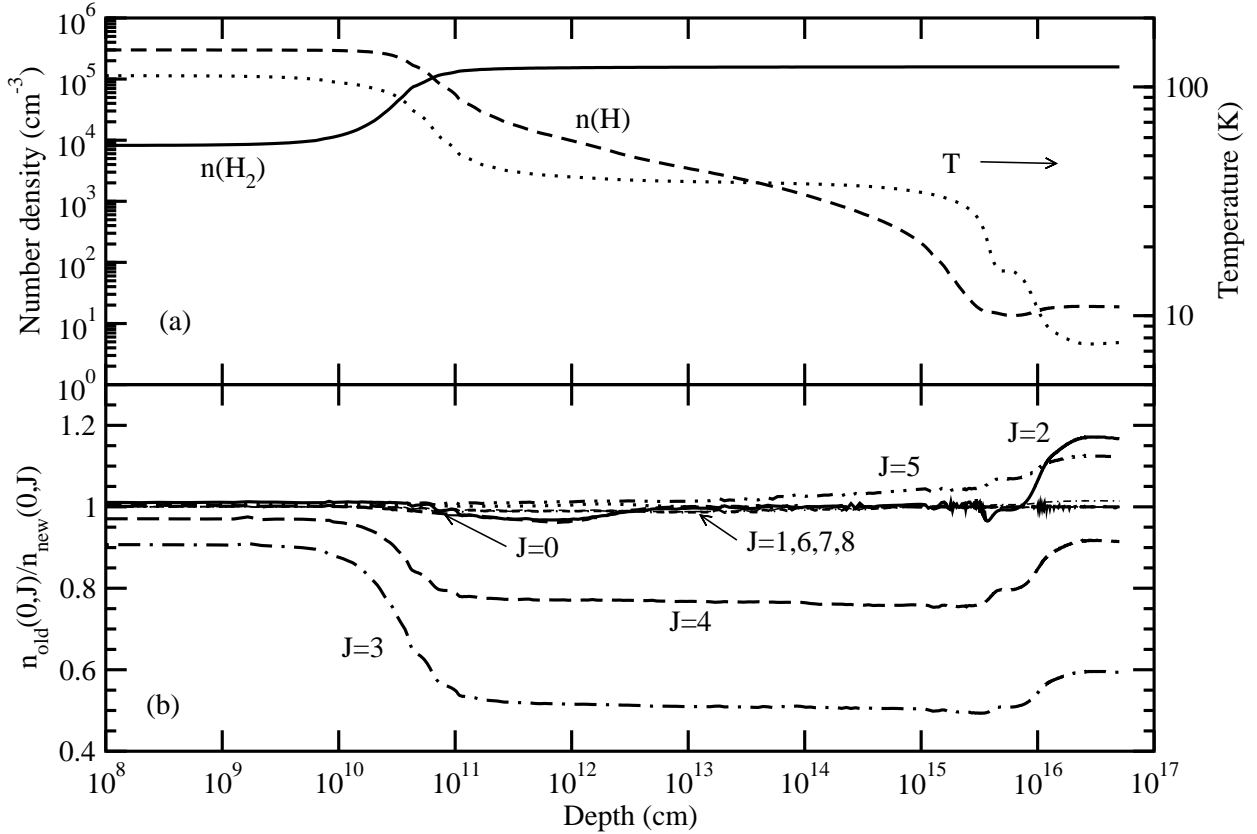} \caption{(a) Total H and H$_2$ abundances and
the gas temperature as a function of depth into a model molecular
photodissociation region. See text for model parameters.
(b) Effect of new H$_2$-H$_2$ rate
coefficients on the H$_2$ rotational populations in the model 
molecular photodissociation 
region. Ratio of $n_{\rm H_2}$($v=0,J=0-8$) obtained with the standard \citet{leb99}
rate coefficient fits (based on Flower 1998) to those obtained 
by replacing the H$_2$-H$_2$ rate coefficients
with the current calculations. See text for details.
\label{fig5}}
\end{figure}
\clearpage

\begin{table}
\caption{H$_2$--H$_2$ interaction energy (in hartrees) for
[$\theta_1$, $\theta_2$, $\phi_{12}$]=[0, 0, 0] (upper entries)
and [$\theta_1$, $\theta_2$, $\phi_{12}$]=[$\pi/2$, $\pi/2$, 0]
(lower entries) configurations, respectively.}

\vspace{.2cm}
\begin{tabular}{ccc}
\hline\hline

$R$(Bohr) & Flower (1998) & Diep \& Johnson (2000) \\

\hline

3.0& 0.06546& 0.061703 \\

3.5&0.02836& 0.027103 \\

4.0&0.01168& 0.011378 \\

4.5&0.004593& 0.004525 \\

5.0& 0.001687 & 0.001660\\
\hline
3.0& 0.03906  & 0.038403\\

3.5& 0.01679& 0.016568\\

4.0& 0.006942 &  0.006585 \\

4.5& 0.002761&  0.002343  \\

5.0&0.001021&   0.000668 \\
\hline
\end{tabular}
\label{tab1}
\end{table}

\begin{table*}
\begin{center}
\caption{Rotational deexcitation rate coefficients as a function of
temperature for H$_2$($v_1=0,J_1$) due to collisions of H$_2$($v_2=0,J_2=0$ or 1).}
\begin{tabular}{cccccccccc}
\tableline
\tableline
\multicolumn{1}{c}{$v_1$} & \multicolumn{1}{c}{$J_1$} &
\multicolumn{1}{c}{$v'_1$} & \multicolumn{1}{c}{$J'_1$} &
\multicolumn{1}{c}{$v_2$} & \multicolumn{1}{c}{$J_2$} &
\multicolumn{1}{c}{$v'_2$} & \multicolumn{1}{c}{$J'_2$} &
\multicolumn{1}{c}{$T$ (K)} & \multicolumn{1}{c}{$k$ (cm$^3$ s$^{-1}$)} \\
\tableline
0 & 2 & 0 &  0 &  0 & 0 & 0 & 0 & 2.0 &  8.365(-13)$^a$ \\ 
0 & 2 & 0 &  0 &  0 & 0 & 0 & 0 & 3.0 &  7.416(-13) \\ 
0 & 2 & 0 &  0 &  0 & 0 & 0 & 0 &  4.0 &  6.531(-13) \\
\multicolumn{10}{l}{The remainder of this table is available only on-line as a machine readable table} \\
\tableline
$^a$The notation $A(-B)$ refers to $A\times 10^{-B}$.
\end{tabular}
\end{center}
\label{tab2}
\end{table*}

\end{document}